\documentclass[english,sort&compress]{iopart}
\usepackage[T1]{fontenc}
\usepackage[latin9]{inputenc}
\usepackage{geometry}
\usepackage{amsbsy}
\usepackage{amstext}
\usepackage[numbers]{natbib}
\usepackage{xargs}[2008/03/08]

\makeatletter
\usepackage{iopams}
\usepackage{setstack}

\usepackage{hyperref}

\newcommand{\eqref}[1]{(\ref{#1})}

\makeatother

\usepackage{babel}
\begin{document}

\title[The Algebra of the Pseudo-Observables III]{The Algebra of the Pseudo-Observables III: Transformations and Time
Evolution}

\author{Edoardo Piparo\footnote{A.I.F. Associazione per l'Insegnamento della Fisica - Gruppo Storia della Fisica: \url{http://www.lfns.it/STORIA/index.php/it/chi-siamo}}}

\address{{\large{}Liceo Scientifico Statale ``Archimede'', Viale Regina
Margherita 3, I-98121 Messina, Italy}}

\ead{{\large{}edoardo.piparo@istruzione.it}}
\begin{abstract}
This paper is the third part concluding the introduction of the powerful
algebra of the pseudo-observables. In this article will be dealt how
to treat the time evolution, and, more in general, the transformations,
in the framework of the new theory. It will be shown that this requires
only minor changes with respect to the Dirac-Jordan transformation
theory that can be found in almost any textbook, with some more care
about the treatment of the continuous limit. A remarkable difference
is also in the introduction of the time reversal, which gives the
opportunity to a deeper insight about the Hermitian transposition.
In the conclusions, we will examine better the relationship between
time evolution and measurement, a very problematic aspect in the framework
of the Copenhagen interpretation (and in many others ones). Finally
we will present a list of compelling reasons for which the physics
community would seriously have to evaluate a switching to the new
formalism.
\end{abstract}



\maketitle
\global\long\def\ket#1{|#1\rangle}

\global\long\def\bra#1{\langle#1|}

\global\long\def\inprod#1#2{\left\langle #1\mid#2\right\rangle }

\global\long\def\ideq{:=}

\global\long\def\imp#1{#1_{\mathrm{I}}}

\global\long\def\rp#1{#1_{\mathrm{R}}}

\newcommandx\der[3][usedefault, addprefix=\global, 1=]{\frac{\rmd^{#1}#2}{\rmd#3^{#1}}}

\newcommandx\pder[3][usedefault, addprefix=\global, 1=]{\frac{\partial^{#1}#2}{\partial#3^{#1}}}

\newcommandx\sinc[1][usedefault, addprefix=\global, 1=]{\mbox{sinc}#1}

\newcommandx\sspan[1][usedefault, addprefix=\global, 1=]{\mbox{span}#1}

\global\long\def\nepe{\rme}

\section{Introduction}

This third paper completes the reformulation of Quantum Mechanics
in the framework of the algebra of the pseudo-observables. After the
introduction of the algebra in \citep{Piparo-I}, referred from now
on as the first paper, in order to show that quantum mechanics is
the unique minimal description of physical reality and the presentation
of the hierarchical structure of the observer networks, we fronted
the measurement problem in \citep{Piparo-II}, referred from now on
as the second paper, giving, in our opinion, a solution to it and
bringing a deeper insight on what to mean for ``physical reality''.

We will deal how to treat the time evolution, and, more in general,
the transformations, in the framework of the algebra of the pseudo-observables.

From an historical point of view, the formulation of the quantum mechanical
equations of motion wasn't much linear. In 1925-26, on the one side,
Werner Heisenberg\citep{Heisenberg1925}, Max Born and Pascual Jordan\citep{M.Born1925,M.Born1925a}
gave rise to \emph{matrix mechanics}, in the framework of which time
evolution is described by the \emph{Heisenberg equation}; on the other
side, Erwin Schrödinger\citep{Schroedinger1926d}, carrying forward
the ideas of de Broglie, created wave mechanics, in which time evolution
is described by the well-known \emph{Schrödinger equation}. The two
approaches found a reconciliation essentially in the 1930 classic
\emph{The Principles of Quantum Mechanics}\citep{Dirac1930}, in which
Paul Dirac showed that Schrödinger's and Heisenberg's approaches were
two different representations of the same theory. The investigations
in the field of the quantum statistical mechanics, led also to the
introduction the \emph{density matrix}\citep{Neumann1927}, whose
time evolution is described by means of the \emph{von Neumann equation}.

In this paper we will show as all these equations can be found in
the framework of the algebra of the pseudo-observables, by following
procedures very similar to those found in almost any textbook, according
the spirit of the \emph{Dirac-Jordan transformation theory}\citep{Dirac1927}.
The main difference is that we will work with finite Hilbert spaces
and a discretized time evolution, recovering the continuity by finally
assuming suitable limits.

\section{Transformations of pseudo-observables }

\subsection{Transformations\label{subsec:Transformations}}

The study of the \emph{physical phenomena} is indeed the more important
aspect of the physical analysis of reality. A physical phenomenon
consists in an evolution process of the physical properties of a system.
It is, therefore, characterized in a natural way through \emph{transformations}.
A transformation is a functional bound between an initial situation
and a final situation. A first hypothesis that it is necessary to
assume about such bound is that it is a biunivocal correspondence
between observables - and more generally pseudo-observables - in the
sense that it must be possible to trace the effects (final situation)
to the causes (initial situation). By virtue of this \emph{reversibility
principle}, therefore, a transformation must be an \emph{invertible
map} among pseudo-observables. By indicating with $\tau$ such a map,
it must transform observables into observables and so, if $O$ is
a generic observable, it must result:
\begin{equation}
O^{\dagger}=O\;\Rightarrow\;\left(\tau\left(O\right)\right)^{\dagger}=\tau\left(O\right)\text{ .}\label{eq:Trasf_Hermitiana}
\end{equation}
The map must also preserve the links between pseudo-observables, it
has, therefore, to be a \emph{ring automorphism}. This implies that,
for two given pseudo-observables $A$ and $B$, $\tau$ must be additive:
\begin{equation}
\tau\left(A+B\right)=\tau\left(A\right)+\tau\left(B\right)\label{eq:Trasf_Somma}
\end{equation}
and must preserve the multiplication relationship:
\begin{equation}
\tau\left(AB\right)=\tau\left(A\right)\tau\left(B\right)\text{ .}\label{eq:Trasf_Prod}
\end{equation}
We will make the further assumption that constants (real or complex)
are \emph{invariant} under transformations, i.e. that, if $\gamma$
is a whatever complex constant, it results:
\begin{equation}
\tau\left(\gamma\right)=\gamma\text{ .}\label{eq:Invarianza_Costanti}
\end{equation}
By these two fundamental requirements, the two important lemmas immediately
derive:
\begin{enumerate}
\item If $P$ is a whatever pseudo-observable, of real part $\rp P$ and
imaginary part $\imp P$, one has:
\[
\tau(P^{\dagger})=\tau(\rp P-i\,\imp P)=\left(\tau(\rp P)+i\,\tau\left(\imp P\right)\right)^{\dagger}=\left(\tau\left(P\right)\right)^{\dagger}\text{ .}
\]
\item Transformations are \emph{linear mappings}, so, if $\gamma_{1}$ and
$\gamma_{2}$ are two complex constants and $P_{1}$ and $P_{2}$
are two pseudo-observables, one has:
\[
\tau\left(\gamma_{1}\,P_{1}+\gamma_{2}\,P_{2}\right)=\gamma_{1}\,\tau\left(P_{1}\right)+\gamma_{2}\,\tau\left(P_{2}\right)\text{ .}
\]
\end{enumerate}
We will show now how these properties full characterize the transformations.

We start analyzing what happens applying a transformation $\tau$
to the projectors. The following theorems hold:
\begin{enumerate}
\item By applying a transformation to a projectors $I$, another projector
is obtained. In fact, $\tau\left(I\right)$, by virtue of hypothesis,
is an observable and, according to the \eqref{eq:Trasf_Prod}, one
has:
\[
\left(\tau\left(I\right)\right)^{2}=\tau\left(I^{2}\right)=\tau\left(I\right)
\]
having made use of the property (2) in the first paper, that
justifies also the thesis.
\item By applying a transformation to two\emph{ mutually exclusive} projectors
$I_{1}$ and $I_{2}$, one obtains two \emph{mutually exclusive projectors}:
\[
I_{1}I_{2}=0\;\Rightarrow\;\tau\left(I_{1}\right)\tau\left(I_{2}\right)=\tau\left(I_{1}I_{2}\right)=\tau\left(0\right)=0\text{ .}
\]
\item By applying a transformation to a set $\left\{ I_{j}\right\} $ of
projectors for which it holds the closure relation (7) in the
first paper, one obtains a set of projectors satisfying the same closure
relation:
\[
\sum_{j}I_{j}=1\;\Rightarrow\;\sum_{j}\tau\left(I_{j}\right)=\tau\left(\sum_{j}I_{j}\right)=\tau\left(1\right)=1\text{ .}
\]
\end{enumerate}
One can, therefore, infer that \textbf{by applying a transformation
to a projector basis one obtains a new projector basis}. We, finally,
prove that by applying a transformation to an \emph{elementary} projector,
another \emph{elementary} projector is obtained. Let, in fact, $I$
be an elementary projector and suppose by absurd $\tau\left(I\right)$
not to be elementary, that is there exist two non-null mutually exclusive
projectors $J_{1}$ and $J_{2}$ such that:
\begin{equation}
\tau\left(I\right)=J_{1}+J_{2}\text{ .}\label{eq:Dim_Trasf_Ind_El}
\end{equation}
By applying to both sides of the \eqref{eq:Dim_Trasf_Ind_El} the
inverse transformation $\tau^{-1}$, then, one would have:
\[
I=\tau^{-1}\left(J_{1}\right)+\tau^{-1}\left(J_{2}\right)
\]
that it would be an absurd, since, for what previously proved, this
would imply that $I$ is equal to the sum of two non-null mutually
exclusive projectors, because transformations, according their fundamental
properties, associate non-null pseudo-observables to non-null pseudo-observables.

We will, now, analyze the effects of a transformation upon a dyad
basis $\left\{ \Gamma_{jk}\right\} $, proving that the set $\left\{ \tau\left(\Gamma_{jk}\right)\right\} $
is a new dyad basis. To this end, we write every dyad $\Gamma_{jk}$
of the basis $\left\{ \Gamma_{jk}\right\} $ as a dyadic form relative
to the pair $\left(I_{j},I_{k}\right)$ of elementary projectors,
belonging to an elementary projector basis $\left\{ I_{j}\right\} $,
having the core $C_{jk}$:
\begin{equation}
\Gamma_{jk}=I_{j}C_{jk}I_{k}\text{ .}\label{eq:Forma_Diadica_Diade}
\end{equation}
By applying the transformation $\tau$ to both sides of \eqref{eq:Forma_Diadica_Diade},
one has:
\[
\tau\left(\Gamma_{jk}\right)=\tau\left(I_{j}\right)\tau\left(C_{jk}\right)\tau\left(I_{k}\right)\text{ .}
\]
Remembering what has been above proved about the application of a
transformation to an elementary projector basis, one concludes that
also $\tau\left(\Gamma_{jk}\right)$ are dyadic forms, which forms
a basis. Observing, besides, that it results:
\begin{enumerate}
\item $\tau\left(\Gamma_{jj}\right)=\tau\left(I_{j}\right)$
\item $\tau\left(\Gamma_{jk}\right)^{\dagger}=\tau\left(\Gamma_{jk}^{\dagger}\right)=\tau\left(\Gamma_{kj}\right)$
\item $\tau\left(\Gamma_{jl}\right)\tau\left(\Gamma_{l'k}\right)=\tau\left(\Gamma_{jl}\Gamma_{l'k}\right)=\delta_{l,l'}\,\tau\left(\Gamma_{jk}\right)$
\end{enumerate}
it is concluded that also the set $\left\{ \tau\left(\Gamma_{jk}\right)\right\} $
is a dyad basis, i.e. \textbf{by applying a transformation to a dyad
basis a new dyad basis is obtained}. For what stated in subsection
4.3 in the first paper, it will therefore have to exist a change
of basis unitary pseudo-observable $W$ such that it results:
\begin{equation}
\tau\left(\Gamma_{jk}\right)=W\,\Gamma_{jk}W^{\dagger}\text{ .}\label{eq:Eq_Trasf_Diadi}
\end{equation}
We will say that the transformation is \emph{induced} by the unitary
pseudo-observable $W$ and that $W$ is the unitary pseudo-observable
\emph{associated} to the transformation.

Let, now, $P$ be a whatever element of the space $\mathfrak{P}$
of the pseudo-observables. One can decompose $P$ according the dyad
basis $\left\{ \Gamma_{jk}\right\} $, as in equation (48) in
the first paper, obtaining:
\[
P=\sum_{j,k}\varpi_{jk}\,\Gamma_{jk}
\]
where the components $\varpi_{jk}$ are suitable complex constants.
By applying to both sides of this relation the transformation $\tau$
and exploiting its linearity, the \eqref{eq:Eq_Trasf_Diadi} and the
distributivity of the product of pseudo-observables over addition,
one obtains the following expression for the transformation equation
in terms of the associated unitary pseudo-observable:
\begin{eqnarray}
\tau\left(P\right) & = & \sum_{j,k}\varpi_{jk}\,\tau\left(\Gamma_{jk}\right)=\sum_{j,k}\varpi_{jk}\,W\,\Gamma_{jk}W^{\dagger}=\nonumber \\
 & = & W\left(\sum_{j,k}\varpi_{jk}\,\Gamma_{jk}\right)W^{\dagger}=WPW^{\dagger}\text{ .}\label{eq:Forma_Unit_Trasf}
\end{eqnarray}

Since $W$ is an unitary pseudo-observable, its real part $\rp W$
is compatible with its imaginary part$\imp W$. In fact, according
to equations (21) and (66), both of them in the first
paper, one has:
\[
\left[\rp W,\imp W\right]=\left[\frac{W+W^{\dagger}}{2},\frac{W-W^{\dagger}}{2i}\right]=\frac{W^{\dagger}W-WW^{\dagger}}{2i}=\frac{1-1}{2i}=0\text{ .}
\]
The real and the imaginary part of $W$ will be therefore allowed
to be expressed as linear combinations of projectors of the same basis
$\left\{ I_{j}\right\} $:
\begin{equation}
\rp W=\sum_{j}\alpha_{j}\,I_{j}\qquad\mbox{and}\qquad\imp W=\sum_{j}\beta_{j}\,I_{j}\label{eq:Parti_RI_W}
\end{equation}
where $\alpha_{j}$ and $\beta_{j}$ are suitable real coefficients.
By making use of the \eqref{eq:Parti_RI_W}, one finds, therefore,
the following expression for the unitary pseudo-observable $W$ associated
to the transformation:
\begin{equation}
W=\sum_{j}\left(\alpha_{j}+i\beta_{j}\right)I_{j}\text{ .}\label{eq:Forma_Complessa_W}
\end{equation}
By substitution of this expression in equation (66) in the first
paper, one obtains the following relation among the coefficients:
\begin{equation}
\alpha_{j}^{2}+\beta_{j}^{2}=1\label{eq:Condizione_Unit}
\end{equation}
valid for each index $j$. The \eqref{eq:Condizione_Unit} implies
that, for each index $j$, it exists an angle $\vartheta_{j}$, that
can be supposed to be, for instance, between $-\pi$ and $\pi$, such
that it results:
\begin{equation}
\begin{array}{l}
\alpha_{j}=\cos\vartheta_{j}\text{ ,}\\
\beta_{j}=\sin\vartheta_{j}\text{ .}
\end{array}\label{eq:Forma_Trig_Coeff_S}
\end{equation}
By substitution of these relations in the \eqref{eq:Forma_Complessa_W},
finally one has:
\begin{equation}
W=\sum_{j}\left(\cos\vartheta_{j}+i\sin\vartheta_{j}\right)I_{j}=\sum_{j}\nepe^{i\,\theta_{j}}I_{j}\label{eq:Forma_Trig_W}
\end{equation}
having made use of the \emph{Euler's formula}.

By considering, now, the observable:
\begin{equation}
G\ideq\sum_{j}\vartheta_{j}\,I_{j}\label{eq:Def_Generatrice}
\end{equation}
according to the definition of a function of an observable, equation
(12) in the first paper, one has:
\begin{equation}
W=\cos\left(G\right)+i\sin\left(G\right)=\nepe^{iG}\text{ .}\label{eq:Forma_Gen_W}
\end{equation}
We will call $G$ the \emph{generatrix} of the transformation. Note,
however, that it is not uniquely identified by the transformation
itself, since the elements of its spectrum are defined to the less
of $2\pi$ multiples.

Consider, now, an observable $A$, expressed as a linear combination
of the projectors of the associated basis:
\[
A=\sum_{j}a_{j}\,I_{A=a_{j}}\text{ .}
\]
By applying to both side the transformation $\tau$, by virtue of
the property of linearity, one obtains:
\begin{equation}
\tau\left(A\right)=\sum_{j}a_{j}\,\tau\left(I_{A=a_{j}}\right)\label{eq:Scomp_Trasform}
\end{equation}
where, according to what stated above, the set $\left\{ \tau\left(I_{A=a_{j}}\right)\right\} $
of observables is a projector basis. Since the coefficients $a_{j}$
are all distinguished among themselves, by comparison of the \eqref{eq:Scomp_Trasform}
with the equation (8) in the first paper, it is concluded that
$\tau\left(A\right)$ has the same spectrum of $A$ and that besides
it results:
\begin{equation}
\tau\left(I_{A=a_{j}}\right)=I_{\tau\left(A\right)=a_{j}}\text{ .}\label{eq:Trasf_Indicatori}
\end{equation}

Consider, finally, a function $f$, defined according the equation
(14) in the first paper, of a complete set $\left\{ O_{r}\right\} $
of compatible observables, which, decomposed according the elementary
projector basis $\left\{ I_{j}\right\} $ associated to the space
of the observables compatible with them, result given by:
\begin{equation}
O_{r}=\sum_{j}o_{r,j}\,I_{j}\text{ .}\label{eq:Scomp_Ins_Comp_Oss}
\end{equation}
If you indicate with $\boldsymbol{O}$ the $n$-tuple formed by the
elements of the complete set of compatible observables, by applying
to the observable $f\left(\boldsymbol{O}\right)$ the transformation
$\tau$, by virtue of the linearity property ad of the \eqref{eq:Scomp_Trasform},
one has:
\begin{equation}
\tau\left(f\left(\boldsymbol{O}\right)\right)=\sum_{j}f\left(\boldsymbol{o}_{j}\right)\tau\left(I_{j}\right)=f\left(\tau\left(\boldsymbol{O}\right)\right)\label{eq:Trasform_Funzione}
\end{equation}
where it was put:
\begin{equation}
\tau\left(\boldsymbol{O}\right)\ideq\left(\tau\left(O_{1}\right),\,\ldots,\,\tau\left(O_{r}\right),\,\ldots\right)\label{eq:Trasf_Lista_Oss}
\end{equation}
i.e. \textbf{by applying a transformation to a function of compatible
observables it is obtained as a result the function of the observables
obtained by applying the transformation to the starting observables}.

\subsection{Transformation invariance\label{subsec:Transformation-invariance}}

It is, now, interesting to characterize the main \emph{invariants}
under a transformation. In this context, an invariant is an observable,
or more generally a pseudo-observable, or a value, real or complex,
associated to one or more observables or pseudo-observables, that
does not change by applying the transformation.

We will begin by demonstrating that \textbf{the observables invariant
under a transformation $\tau$, induced by the unitary pseudo-observable
$W$, are all and only those compatible with the generatrix $G$ of
the transformation}. Let, in fact, $A$ be an observable compatible
with $G$ and so commuting with it:
\[
\left[A,G\right]=0
\]
Since if two observables commute also each function of the one commutes
with every function of the other, the above relation implies:
\[
\left[A,\cos\left(G\right)\right]=\left[A,\sin\left(G\right)\right]=0
\]
and therefore:
\[
\left[A,W\right]=\left[A,\cos\left(G\right)+i\sin\left(G\right)\right]=\left[A,\cos\left(G\right)\right]+i\left[A,\sin\left(G\right)\right]=0\text{ .}
\]
By the last relation it follows:
\[
AW=WA
\]
by which, by multiplying both sides by $W^{\dagger}$ to the right
and exploiting the unitarity of $W$, one obtains:
\[
A=WAW^{\dagger}=\tau\left(A\right)
\]
which proves that each observable compatible with $G$ is also an
invariant under the transformation. If, conversely, an observable
$A$ is an invariant under the transformation $\tau$, it will be:
\[
A=\tau\left(A\right)=WAW^{\dagger}\text{ .}
\]
By multiplying both sides of this relation to the right by $W$ and
exploiting the unitarity of this last, one has:
\begin{equation}
AW=WA\label{eq:Commutazione_con_W}
\end{equation}
that is the observable $A$ commutes with $W$. By transposition of
both sides of the \eqref{eq:Commutazione_con_W}, then one has that
$A$ also commutes with $W^{\dagger}$. It is so concluded that $A$
is compatible both with the real part and the imaginary part of $W$.
There will therefore be an elementary projector basis with respect
to which it is possible express, as linear combinations of the elements
of the basis, both the observable $A$ and $W$, and so, by virtue
of the definition \eqref{eq:Def_Generatrice}, also the generatrix
$G$. The observables $A$ and $G$ will be, therefore, compatible

We will, now, prove that the \textbf{trace of an observable}, as defined
in subsection 2.2 in the second paper,\textbf{ is an invariant
properties of the transformation}. To this end it suffices to observe
that a transformation, as showed at the end of subsection \eqref{subsec:Transformations},
does not alter nor the observable spectrum, nor the multiplicity of
the relative terms and therefore, according to definition (9)
in the second paper, does not change the trace. More in general, for
a whatever pseudo-observable $P$, by virtue of the third property
of the trace, one has:
\begin{equation}
\tr\left(\tau\left(P\right)\right)=\tr\left(WPW^{\dagger}\right)=\tr\left(PW^{\dagger}W\right)=\tr\left(P\right)\label{eq:Invarianza_Traccia}
\end{equation}
where it was also exploited the unitarity of the pseudo-observable
$W$ associated to the transformation $\tau$. 

A further important consequence of the \eqref{eq:Invarianza_Traccia},
by virtue of the definition (23) in the second paper, of the
relation \eqref{eq:Trasf_Prod} and of the first lemma about transformations,
is that also \textbf{inner products are invariant under transformations}.
By considering, in fact, two whatever pseudo-observables $X$ and
$Y$, one has:
\begin{eqnarray}
\inprod{\tau\left(X\right)}{\tau\left(Y\right)} & = & \tr\left(\tau\left(X\right)^{\dagger}\,\tau\left(Y\right)\right)=\tr\left(\tau\left(X^{\dagger}Y\right)\right)=\nonumber \\
 & = & \tr\left(X^{\dagger}Y\right)=\inprod XY\text{ .}\label{eq:Invarianza_Prod_Scalare}
\end{eqnarray}
It should be, however, pointed out that one is \textbf{not allowed
}to infer by such property that the expectation value of an observable
or of a pseudo-observable is invariant under a transformation, despite
the fact that, according to the equation (75) in the second
paper, this can be expressed as the inner product between the pseudo-observable
and the density observable. The density observable is, in fact, somewhat
peculiar, being built substantially according an \textit{a posteriori}
process, in order to summarize the statistical properties of the possible
measurements relative to a given system in a certain state, in the
spirit of the \emph{Bayesian inference}\citep{Fuchs2010,Fuchs2017}.
In introducing the concept of transformation, it was implicitly adopted
the \emph{Heisenberg picture}, i.e. it was assumed that \textbf{the
transformations change the observables while leaving unaltered the
states}. In such a picture one has, therefore, to assume that \textbf{the
density observable relative to the transformed observables is the
same of that relative to the starting ones}. The expectation value
of a pseudo-observable $P$, to which is applied a transformation
$\tau$, will therefore change, on the basis of the equation (75)
in the second paper, according to the law:
\begin{equation}
\left\langle P\right\rangle =\inprod DP\overset{\tau}{\longmapsto}\left\langle \tau\left(P\right)\right\rangle =\inprod D{\tau\left(P\right)}\label{eq:Trasf_Valore_Medio}
\end{equation}

According to the \eqref{eq:Trasf_Valore_Medio} and exploiting the
invariance of the inner product, equation \eqref{eq:Invarianza_Prod_Scalare},
besides, one has:
\begin{equation}
\left\langle \tau\left(P\right)\right\rangle =\inprod D{\tau\left(P\right)}=\inprod{\tau\left(\tau^{-1}\left(D\right)\right)}{\tau\left(P\right)}=\inprod{\tau^{-1}\left(D\right)}P\label{eq:Schema_Schroedinger}
\end{equation}
On the basis of such relation, it is therefore inferred that\textbf{
the expectation values of observables and pseudo-observables are invariant
for those states for which the density observable is invariant under
the transformation, i.e.}, according to what proved at the beginning
of this subsection, \textbf{for those ones for which $D$ is compatible
with the generatrix $G$ of the transformation}.

Equation \eqref{eq:Schema_Schroedinger}, shows as the expectation
value after applying a transformation can be calculated as the starting
one by making use of a suitably transformed density observable, so
justifying, as an alternative to the Heisenberg picture, the well-known
\emph{Schrödinger picture}, in which \textbf{the observables remain
unaltered while to evolve are the states, described in terms of the
density $D$}. In such a picture, as a result of a transformation
$\tau$, the system state changes, passing from the situation described
by the density $D$ to the one described by the density $D'=\tau^{-1}\left(D\right)$,
that is:

\begin{equation}
D\overset{\tau}{\longmapsto}D'=\tau^{-1}\left(D\right)\text{ .}\label{eq:Trasf_Densita}
\end{equation}

If $\Psi_{j}$ is the state vector associated to the pure state of
index $j$, the expectation value a pseudo-observable $P$, according
equation (86) in the second paper, is given by:

\[
\left\langle P\right\rangle _{j}=\inprod{\Psi_{j}}{P\Psi_{j}}\text{ .}
\]

After applying the transformation $\tau$, according to \eqref{eq:Forma_Unit_Trasf},
this will change becoming:

\begin{equation}
\left\langle \tau\left(P\right)\right\rangle _{j}=\inprod{\Psi_{j}}{\tau\left(P\right)\Psi_{j}}=\inprod{W^{\dagger}\Psi_{j}}{P\,W^{\dagger}\Psi_{j}}\text{ .}\label{eq:Val_Medio_Trasf}
\end{equation}

It is easily verified that, due to unitarity of $W$, the set $\left\{ W^{\dagger}\Psi_{j}\right\} $
has all the properties of a\emph{ set of state vectors}, as defined
in section 2 of the second paper, relative to the anti-transformed
dyad basis. This implies that one may consider $W^{\dagger}\Psi_{j}$
as the transformed state vector associated to the pure state of index
$j$, that is:
\begin{equation}
\Psi_{j}\overset{\tau}{\longmapsto}{\Psi'}_{j}=W^{\dagger}\Psi_{j}\label{eq:Trasf_Vettori_Stato}
\end{equation}
that is, therefore, the\textbf{ law expressing the transformation
of the state vectors in the Schrödinger picture}.

\subsection{Translations\label{subsec:Translations}}

Consider an observable $A$ decomposed, according to the (8)
in the first paper, as a linear combination of the projectors of the
associated basis:
\[
A=\sum_{j}a_{j}\,I_{A=a_{j}}
\]
clearly independent on the order of the terms in the summation. We
will agree in sorting the spectrum elements in ascending order, such
that it results: $j>k\;\Rightarrow\;a_{j}>a_{k}$.

By indicating with $\epsilon$ an arbitrarily chosen positive real
constant, the sequence $\left\{ j\epsilon\right\} $ is monotonically
increasing and therefore, by a suitable restriction of the number
of its terms, can be put in a biunivocal correspondence with that
of the spectrum elements of $A$, through the mapping:
\begin{equation}
f\,:\,j\epsilon\rightarrow a_{j}\label{eq:Funz_Lineariz}
\end{equation}
According to the definition given in the first paper, equation (11),
we will be therefore allowed to think $A$ as a function of the observable
$A_{\mathrm{L}}$, defined by the equation:
\begin{equation}
A_{\mathrm{L}}\ideq\sum_{j}j\epsilon\,I_{A=a_{j}}\label{eq:Osserv_Lineariz}
\end{equation}
that is:
\[
A=f\left(A_{\mathrm{L}}\right)
\]
where $f$ is the function defined in \eqref{eq:Funz_Lineariz}.

We will call an observable whose spectrum is of the form of that of
$A_{\mathrm{L}}$ a \textbf{\emph{linear spectrum observable}}. The
constant $\epsilon$ is the \emph{resolution} of the observable, that
is the minimal variation possible for the quantity. The arbitrarily
in the choose of its value reflects the arbitrariness in choosing
the units of measurement.

Let $Q$ be a linear spectrum observable with resolution $\epsilon$.
We will suppose, besides, that the number of the elements of the spectrum
of $Q$ is very large, such that one is legitimated to assume it equal
to $2n\approx2n+1$, where $n$ is a natural number much greater than
$1$. In other words the number of the elements of the spectrum is
a \emph{potential infinite}. This assumption is justified by mathematical
convenience reasons, in order to avoid several troubles with actual
infinities, and by the fact that the set of the effective outcomes
obtained in all the measurement sessions is necessarily finite. According
to the \eqref{eq:Osserv_Lineariz}, we will then be able to decompose
$Q$ as:
\begin{equation}
Q=\sum_{j}j\epsilon\,I_{j}\label{eq:Scomp_Q}
\end{equation}
where $\left\{ I_{j}\right\} $ is the projector basis associated
to the observable and the index $j$ assumes all the integer values
between $-n$ and $n$.

We will assume that all the \emph{generalized coordinates}, required
to describe a physical system at any given time instant, are linear
spectrum observables (\textbf{\emph{homogeneity hypothesis}}).

By indicating with $\delta$ a real constant, we will introduce, with
the appropriate hypotheses, the \emph{translation} of \emph{displacement}
$\delta$ of the observable $Q$, as that transformation $\tau_{\delta}$
that applied to the observable makes it result:
\begin{equation}
\tau_{\delta}\left(Q\right)=Q+\delta\label{eq:Def_Traslazione}
\end{equation}
in which the sum in the right-hand side is to be meant in a manner
specified below.

To this end, we firstly decompose the displacement $\delta$ in the
form:
\begin{equation}
\delta=\xi+s\epsilon\label{eq:Scomp_Shift}
\end{equation}
where $s$ is an integer and $\xi$ a real constant between $0$ and
$\epsilon$:
\begin{equation}
0\leq\xi<\varepsilon\text{ .}\label{eq:Limit_Csi}
\end{equation}
It should be observed that such decomposition is unique for each value
of $\delta$.

By substituting the \eqref{eq:Scomp_Q} and the \eqref{eq:Scomp_Shift}
into the \eqref{eq:Def_Traslazione} and exploiting the linearity
of the transformation and the closure relation of the projectors,
equation (7) in the first paper, therefore, it must result:
\begin{eqnarray}
\sum_{j}j\epsilon\,\tau_{\delta}\left(I_{j}\right) & = & \sum_{j}\left(j\epsilon\right)I_{j}+\left(\xi+r\epsilon\right)\sum_{j}I_{j}=\nonumber \\
 & = & \sum_{j}\left(\xi+\left(j+r\right)\epsilon\right)I_{j}=\sum_{j}\left(\xi+j\epsilon\right)I_{j-r}\text{ .}\label{eq:Trasf_Trasl}
\end{eqnarray}
Since, as seen in subsection \ref{subsec:Translations}, the transformed
observable has the same spectrum as the starting one, equation \eqref{eq:Trasf_Trasl}
can be satisfied only by admitting that $\xi=0$, that the sum in
the right-hand side of the \eqref{eq:Def_Traslazione} is modulo $2n\epsilon$
and that the difference $j-r$ is modulo $2n$.

\subsection{Conjugate momenta\label{subsec:Conjugate-momenta}}

If $S=e^{iG}$ is the unitary pseudo-observable associated to the
\emph{minimal translation} ($\delta=\epsilon$) whose generatrix is
$G$, it's easy to verify that the translation of displacement $s\epsilon$
is induced by the unitary pseudo-observable $S^{s}=e^{isG}$, whose
generatrix is $sG$. By virtue of this direct proportionality between
the displacement and the generatrix, one it is allowed to assume that
it results:
\begin{equation}
G\ideq\frac{\epsilon}{\hbar}P\label{eq:Def_Momento}
\end{equation}
where the observable $P$ is the \emph{conjugate momentum} of $Q$
and the \emph{reduced Planck constant }$\hbar$ is introduced for
reasons of choice of the measurement units. The unitary pseudo-observable
associated to the minimal translation can so be written in the form:
\begin{equation}
S=\nepe^{i\frac{\epsilon}{\hbar}P}\label{eq:Trasl_Minima_vs_P}
\end{equation}
and therefore thought as a function of the conjugate momentum. Due
to the periodicity of the complex exponential function, by adding
to the momentum $P$ a whatever multiple of $2\pi\hbar/\epsilon$
the pseudo-observable $S$ does not change. The terms of the spectrum
of the momentum $P$ will be, therefore, distributed in an interval
of diameter of $2\pi\hbar/\epsilon$ and one will be allowed to assume
that the generic eigenvalue $p_{k}$ of $P$ satisfies the limitations:
\begin{equation}
-\frac{\pi\hbar}{\epsilon}\le p_{k}\le\frac{\pi\hbar}{\epsilon}\text{ .}\label{eq:Limit_Spettro_P}
\end{equation}
The fact that the sum in the right-hand side of the \eqref{eq:Def_Traslazione}
is modulo $2n\epsilon$, implies that it has to be:
\[
\tau_{2n\epsilon}\left(Q\right)=Q
\]
and therefore:
\begin{equation}
S^{2n}=1\label{eq:Cond_Periodiche}
\end{equation}
that is:
\begin{equation}
2n\frac{\epsilon}{\hbar}p_{k}=2k\pi\;\Rightarrow p_{k}=k\frac{\pi\hbar}{n\epsilon}\qquad\mbox{with}\qquad k=-n,\cdots,n\label{eq:Valori_Momento}
\end{equation}
where it was made use of the limitations \eqref{eq:Limit_Spettro_P}
and that $n$ is much greater than $1$. From this follows that also
the momentum $P$ is a linear spectrum observable, with $2n$ elements
in its spectrum, and, therefore, can be written in the form:
\begin{equation}
P=\sum_{k}k\frac{\pi\hbar}{n\epsilon}\,\tilde{I_{k}}\label{eq:Scomp_P}
\end{equation}
where $\left\{ \tilde{I_{k}}\right\} $ is the projector basis associated
to $P$.

In order to better specify the relationship between the observables
$Q$ and $P$, we substitute the \eqref{eq:Def_Momento} in the \eqref{eq:Forma_Unit_Trasf},
obtaining:
\begin{equation}
\tau_{\epsilon}\left(Q\right)=\nepe^{i\frac{\epsilon}{\hbar}P}Q\nepe^{-i\frac{\epsilon}{\hbar}P}=Q+\epsilon\label{eq:Trasl_Minima}
\end{equation}
where, due to the hypothesis that is $n\gg1$, one can neglect that
the sum in the right-hand side would have to be  modulo $2n\epsilon$.
By \eqref{eq:Trasl_Minima}, it immediately follows:
\[
\frac{\nepe^{i\frac{\epsilon}{\hbar}P}Q\nepe^{-i\frac{\epsilon}{\hbar}P}-Q}{\epsilon}=1
\]
that in the limit of the resolution that tends to zero, $\epsilon\rightarrow0^{+}$,
gives the well-known \emph{canonical commutation relation}:
\begin{equation}
\underset{\epsilon\rightarrow0^{+}}{\lim}\,\frac{1}{i\hbar}\left[Q,P\right]=1\label{eq:Parentesi_Poisson}
\end{equation}
which has the structure of a \emph{Poisson bracket} of \emph{canonical
coordinates}.

It is, however, to be observed that the \eqref{eq:Parentesi_Poisson}
is exactly true only in the limits $n\rightarrow\infty$ and $\epsilon\rightarrow0$.
It is, instead, impossible to satisfy for finite value of $n$, as
observed already by Weyl\citep{Weyl1928}. According to the (18)
in the second paper, in fact, it would result: $\tr\left({\left[Q,P\right]}\right)=0$;
whereas, by indicating with $d$ the number of the elementary projectors
in a basis $\left\{ I_{j}\right\} $, that, by the assumptions, is
much greater than $1$, due to the linearity of the trace functional
and according to the (21) in the second paper, for the right-hand
side, one has: $\tr\left(1\right)=\tr\left(\sum_{j}I_{J}\right)=\sum_{j}\tr\left(I_{j}\right)=d\gg1$. 

\section{Time evolution}

\subsection{Time evolution in the Heisenberg picture\label{subsec:Time-evolution-Heisenberg}}

The physical phenomena are describable as evolution processes of observables
while passing of the time. The observation of a physical phenomenon
thus requires the measurement of a certain set of observable at different
times. Let $\tau$ be the \emph{minimum} possible \emph{temporal separation}
between two subsequent measurements. We will assume the \textbf{\emph{hypothesis
of the homogeneity of the time}}, according to which this minimum
time separation is always the same.

In this subsection we will analyze time evolution in the \emph{Heisenberg
picture}.

We start studying the time evolution of an observable characterizing
the system that doesn't explicitly depend on time. This is describable
as a transformation in a new one, having the same spectrum but, usually,
incompatible with the initial observable.

Let $\mathbf{Q}$ be a tuple of generalized coordinates that gives
a complete description of the system and $\mathbf{P}$ the tuple of
the corresponding conjugate momenta. The time evolution of the system
is therefore described by means of a transformation $\theta$ that
makes it pass from the canonical coordinates, $\left(\mathbf{Q}\left(t\right),\mathbf{P}\left(t\right)\right)$,
relative to the time $t$ to those corresponding to the time $t+\tau$,
according to the relations:

\begin{equation}
\begin{array}{l}
\theta\left(\mathbf{Q}\left(t\right)\right)=\mathbf{Q}\left(t+\tau\right)\\
\theta\left(\mathbf{P}\left(t\right)\right)=\mathbf{P}\left(t+\tau\right)\text{ .}
\end{array}\label{eq:Def_Evoluzione}
\end{equation}
By indicating with $U$,\textbf{\emph{ pseudo-observable of minimal
evolution}}, the unitary pseudo-observable associated to such transformation
and calling $G$ its generatrix, similarly as seen for translations,
we will put:
\begin{equation}
G=\frac{\tau}{\hbar}H\label{eq:Def_Hamiltonina}
\end{equation}
where $H$ is the \emph{Hamiltonian observable}.

Consider, now, an observable $O\left(\boldsymbol{Q}\left(t\right),\boldsymbol{P}\left(t\right)\right)$,
given by a function of an observable obtained by summation and multiplication
of the canonical coordinates or of functions of compatible groups
of them. According to the properties of transformations, it will be:
\begin{eqnarray}
O\left(\boldsymbol{Q}\left(t+\tau\right),\boldsymbol{P}\left(t+\tau\right)\right) & = & O\left(\theta\left(\boldsymbol{Q}\left(t\right)\right),\theta\left(\boldsymbol{P}\left(t\right)\right)\right)=\nonumber \\
 & = & \theta\left(O\left(\boldsymbol{Q}\left(t\right),\boldsymbol{P}\left(t\right)\right)\right)\text{ .}\label{eq:Evol_Oss_Ind_t}
\end{eqnarray}
If, for brevity, one puts $O\left(t\right)\ideq O\left(\boldsymbol{Q}\left(t\right),\boldsymbol{P}\left(t\right)\right)$,
according to the \eqref{eq:Forma_Unit_Trasf} and to \eqref{eq:Forma_Gen_W},
therefore, one has:
\begin{equation}
O\left(t+\tau\right)=UO\left(t\right)U^{\dagger}=\nepe^{i\frac{\tau}{\hbar}H}O\left(t\right)\nepe^{-i\frac{\tau}{\hbar}H}\text{ .}\label{eq:Evoluzione_Temporale}
\end{equation}

If the observable $O$ depends also explicitly on time, i.e. it is
of the form:
\[
O\left(\boldsymbol{Q}\left(t\right),\boldsymbol{P}\left(t\right),t\right)\text{ ,}
\]
the evolution may be split into two (simultaneous) steps:
\begin{equation}
O\left(\boldsymbol{Q}\left(t\right),\boldsymbol{P}\left(t\right),t\right)\longmapsto O\left(\boldsymbol{Q}\left(t\right),\boldsymbol{P}\left(t\right),t+\tau\right)\label{eq:Evol_Step1}
\end{equation}
\begin{equation}
O\left(\boldsymbol{Q}\left(t\right),\boldsymbol{P}\left(t\right),t+\tau\right)\longmapsto O\left(\boldsymbol{Q}\left(t+\tau\right),\boldsymbol{P}\left(t+\tau\right),t+\tau\right)\label{eq:Evol_Step2}
\end{equation}
that is in the process \eqref{eq:Evol_Step1}, into which the observable
is made evolve while keeping fixed the canonical coordinates and in
the process \eqref{eq:Evol_Step2}, into which the canonical coordinates
evolve while ignoring the explicit dependence on time of the observable.
According to such splitting and by applying to the second step the
\eqref{eq:Evoluzione_Temporale}, by putting, for brevity, $O\left(t+\tau\right)\ideq O\left(\boldsymbol{Q}\left(t+\tau\right),\boldsymbol{P}\left(t+\tau\right),t+\tau\right)$,
one has:
\begin{equation}
O\left(t+\tau\right)=\nepe^{i\frac{\tau}{\hbar}H\left(t\right)}O\left(\boldsymbol{Q}\left(t\right),\boldsymbol{P}\left(t\right),t+\tau\right)\nepe^{-i\frac{\tau}{\hbar}H\left(t\right)}\text{ .}\label{eq:Evoluzione_Temporale_Gen}
\end{equation}

Treating $\tau$ as an infinitesimal and by a Taylor expansion of
both sides of the \eqref{eq:Evoluzione_Temporale_Gen} to first order
in it, one obtains, by omitting the arguments, the following relation:

\begin{equation}
\der Ot=\frac{1}{i\hbar}\left[O,H\right]+\pder Ot\label{eq:Eq_Heisenberg_Gen}
\end{equation}
that is the general form of the well-known \emph{Heisenberg equation}.

Consider now the \textbf{\emph{temporal abscissa}} observable $T$
of an event. By thinking $T$ as a function of the canonical coordinate,
by virtue of equation \eqref{eq:Evoluzione_Temporale}, one has:
\begin{equation}
e^{i\frac{\tau}{\hbar}H}T\left(t\right)e^{-i\frac{\tau}{\hbar}H}=T\left(t+\tau\right)=T\left(t\right)+\tau\text{ .}\label{eq:Evol_Durata}
\end{equation}
The time evolution transformation $\theta$ then represents for the
temporal abscissa $T$ a translation of minimal (temporal) displacement
$\tau$. 

\subsection{Constants of motion and symmetries\label{subsec:Constants-of-motion}}

The time evolution gives rise to several important \emph{conservation
laws}.

First of all, \textbf{if two observables, that does not depend explicitly
on time, are initially compatible they remain compatible afterward}.
In fact, by indicating with $A(t)$ and $B(t)$ the two observables
at a time $t$ and if it results $\left[A\left(t\right),B\left(t\right)\right]=0$,
for the properties of the transformations, one has:
\begin{equation}
\left[A\left(t+\tau\right),B\left(t+\tau\right)\right]=\left[\theta\left(A\left(t\right)\right),\theta\left(B\left(t\right)\right)\right]=\theta\left(\left[A\left(t\right),B\left(t\right)\right]\right)=0\text{ .}\label{eq:Evol_Commutat}
\end{equation}

Consider, now, a one-parameter group of transformations induced by
unitary pseudo-observables $R_{\xi}$ of the form:
\begin{equation}
R_{\xi}=\nepe^{i\xi F}\label{eq:Trasformazioni_Contatto}
\end{equation}
where $\xi$ is a real parameter and $F$ is an observable. Equation
\eqref{eq:Trasformazioni_Contatto} characterizes a \emph{Lie group},
in the parameter $\xi$, whose generator is the observable $F$. In
the hypothesis that $F$ does not explicitly depend on time, such
group of transformation represents a\emph{ continuous symmetry} for
the system if each transformation leaves unchanged the Hamiltonian
observable:
\begin{equation}
R_{\xi}HR_{\xi}^{\dagger}=H\text{ .}\label{eq:Trasf_Simmetria}
\end{equation}
By a Taylor expansion of the \eqref{eq:Trasformazioni_Contatto} with
respect to $\xi$, treated as in infinitesimal, one obtains:
\[
H+\xi\,i\left[F,H\right]+\frac{\xi^{2}}{2}\,i\left[F,i\left[F,H\right]\right]+\cdots=H
\]
from which it follows:
\begin{equation}
\left[F,H\right]=0\text{ .}\label{eq:Rel_Commut_Simm}
\end{equation}
So we have retrieved, in the formalism of the pseudo-observables,
the well-known \emph{correspondence between the continuous symmetries
and the compatibility among their generators and the Hamiltonian observable}.
According to the Heisenberg equation \eqref{eq:Evoluzione_Temporale},
one therefore has:
\begin{equation}
F(t+\tau)=\nepe^{i\frac{\tau}{\hbar}H}F\left(t\right)\nepe^{-i\frac{\tau}{\hbar}H}=F(t)\label{eq:Costante_Moto}
\end{equation}
that implies the the generator $F$ is a \emph{constant of motion},
as stated, in a more general form, by the \emph{Noether's theorem}\citep{Noether1918}.

As again well-known, equation \eqref{eq:Costante_Moto} implies that
if the Hamiltonian observable does not depend explicitly on time it
is a constant of motion and the terms of its spectrum $\left\{ \varepsilon_{j}\right\} $
give the possible outcomes of energy measurement of the system.

In such hypothesis, consider an observable $O\left(\boldsymbol{Q}\left(t\right),\boldsymbol{P}\left(t\right)\right)$
that also does not depend explicitly on time. The time evolution of
$O$ is given by the \eqref{eq:Evoluzione_Temporale}, from which
it follows:
\begin{equation}
O\left(t\right)=U^{\dagger}O\left(t+\tau\right)U\;\Rightarrow\;O\left(t-\tau\right)=U^{\dagger}O\left(t\right)U\label{eq:Evoluzione_Temp_Inv}
\end{equation}
that describes the\emph{ time reversed evolution} of the observable
$O$. By analyzing such relation, it is argued that, if $Z$ is a
generic pseudo-observable appearing in the \eqref{eq:Evoluzione_Temp_Inv},
the\emph{ time reversal} ($\mbox{T}$) implies the substitutions:
\begin{equation}
\left|\begin{array}{l}
Z\rightarrow Z^{\dagger}\\
\tau\rightarrow-\tau
\end{array}\right.\text{ .}\label{eq:Inversione_Temporale}
\end{equation}
Generalizing such result, it is assumed that time reversal implies
the application of the substitutions \eqref{eq:Inversione_Temporale}
to \textbf{every} pseudo-observable.

By applying the \eqref{eq:Inversione_Temporale} to the pseudo-observable
of minimal evolution $U=\nepe^{i\tau H/\hbar}$, it is argued that
the reversed time evolution is generated by the same Hamiltonian observable
that generates the direct temporal evolution (\emph{reversibility
principle}), provided that this last does not explicitly depend on
time, that is:
\begin{equation}
\mbox{T}\left(H\right)=H\text{ .}\label{eq:Inv_Temp_Hamiltoniana}
\end{equation}

The application of time reversal to the temporal abscissa $T$ of
an event makes, according to the second substitution in the \eqref{eq:Inversione_Temporale},
reverse the sign of each term in its spectrum and therefore:

\begin{equation}
\mbox{T}\left(T\right)=-T\text{ .}\label{eq:Inv_Temp_Tempo}
\end{equation}

Since, besides, the substitutions implied by the \eqref{eq:Inversione_Temporale},
does not affect any generalized coordinate $Q$, one has:
\begin{equation}
\mbox{T}\left(Q\right)=Q\text{ .}\label{eq:Inv_Temp_Q}
\end{equation}

Applying, finally, the time reversal to the unitary pseudo-observable
$S$ associated to the minimal translation of the coordinate $Q,$
since it must remain unaffected, it is argued that time reversal reverses
the signs of the conjugate momenta:
\begin{equation}
\mbox{T}\left(P\right)=-P\text{ .}\label{eq:Inv_Temp_Momento}
\end{equation}

For what above stated, the effect of the time reversal on a generic
observable will be given by:
\begin{equation}
\mbox{T}\left(O\left(\boldsymbol{Q},\boldsymbol{P},t\right)\right)=O\left(\boldsymbol{Q},-\boldsymbol{P},-t\right)\text{ .}\label{eq:Inv_Temp_Var_Din}
\end{equation}

It is important, now, to specify the relationship, given by the first
substitution in the \eqref{eq:Inversione_Temporale}, between time
reversal and transposition. \textbf{Time reversal}, indeed, \textbf{implies
the inversion of the observation order}, as it was already heuristically
stated in subsection 3.2 in the first paper. This association,
therefore, finds here a more rigorous justification.

\subsection{Time evolution in the Schrödinger picture}

The time evolution in the Schrödinger picture can be easily described
by making use of what discussed at the end of subsection \eqref{subsec:Transformation-invariance}.
First of all, it is worth remembering that in this picture the observables
evolves only if they depend \textbf{explicitly} on time. Time evolution
so affects mainly the quantum state, as determined through a Bayesian
inference on the basis of the outcomes of an initial set of compatible
measurements. If the initial state is reputed to be described as a
pure state associated to a state vector $\Psi$, according to the
\eqref{eq:Trasf_Vettori_Stato} and to the definition of the pseudo-observable
of minimal evolution given in subsection \eqref{subsec:Time-evolution-Heisenberg},
one has:
\begin{equation}
\Psi\left(t+\tau\right)=U^{\dagger}\Psi\left(t\right)=\nepe^{-i\frac{\tau}{\hbar}H\left(t\right)}\Psi\left(t\right)\text{ ,}\label{eq:Eq_Schroedinger_discr}
\end{equation}
from which, by treating $\tau$ as an infinitesimal and by a Taylor
expansion of both sides of the \eqref{eq:Eq_Schroedinger_discr} to
first order in it, omitting, for brevity, the time argument, it follows
:
\begin{equation}
i\hbar\der{\Psi}t=H\,\Psi\label{eq:Eq_Schroedinger_SV}
\end{equation}
that is the well-known \emph{Schrödinger equation of motion}.

In the more general case in which the initial state of the physical
system is described, as a mixed state, in terms of the density observable
$D$, according to the \eqref{eq:Trasf_Densita}, one, instead, obtains
the following time evolution equation:
\begin{equation}
D\left(t+\tau\right)=U^{\dagger}D\left(t\right)U=\nepe^{-i\frac{\tau}{\hbar}H\left(t\right)}D\left(t\right)\nepe^{i\frac{\tau}{\hbar}H\left(t\right)}\text{ .}\label{eq:Eq_Liouville_discr}
\end{equation}

By treating again $\tau$ as an infinitesimal and by a Taylor expansion
of both sides of the \eqref{eq:Eq_Liouville_discr} to first order
in it, omitting, for brevity, the time argument, one finally obtains
the well-known \emph{von Neumann equation}:
\begin{equation}
i\hbar\der Dt=\left[H,D\right]\text{ .}\label{eq:Eq_von_Neumann}
\end{equation}

\section{Conclusions}

\subsection{The continuous limit\label{subsec:The-continuous-limit}}

The way followed to derive the results in this article may appear
dissatisfying the Mathematical readers, due to the somewhat coarse
limit procedures used. In this subsection I will try to better justify
the reasons for doing so.

In subsection \ref{subsec:Translations}, the generalized coordinates
were initially introduced as linear spectrum observables, with a discrete
and finite spectrum. We was forced to do this, because of the simplifying
assumptions taken in the first paper. But there is more. A continuous
spectrum implies the physical possibility of resolve two different
eigenvalues, however close they can be. This corresponds to the possibility
of performing an exact measurement of a generalized coordinate $Q$,
that is a measurement whose standard deviation is $0$, but, in force
of the Heisenberg uncertainty relations and of the canonical commutation
relation \ref{eq:Parentesi_Poisson}, this would involve an actual
infinite deviation in the conjugate momentum, that would reflect itself
in an actual infinite amount of energy to be used in the measurement
process. Since this is impossible, this implies that it is also impossible
to experimentally prove the continuity of the spectrum of a generalized
coordinate. It is also clearly impossible to experimentally prove
the infinity of the spectrum and also its boundlessness. This concepts
exist only in the Mathematics Realm, since it is true what Kronecker,
as quoted by Weber in 1893, said:
\begin{quotation}
``Die ganzen Zahlen hat der liebe Gott gemacht, alles andere ist
Menschenwerk''

{[}``God made the integers, all else is the work of man''{]}
\end{quotation}
but also actual infinities, with all their well-known paradoxes, cannot
find place in the Physics World.

This may appear again as an excess of ``empiricism'', but there
is other. The validity of the canonical commutation relation requires
the boundless of at least one of a pair of canonical coordinates.
This boundlessness cannot be approximate, not even as a result of
a limit, since the argument of Weyl, readjusted for pseudo-observables
in subsection \ref{subsec:Conjugate-momenta}, proves that it isn't
even approximately true for finite and bounded canonical coordinates.
At this regard, it is not to be forgotten that the whole canonical
formalism relies on the hypothesis of the continuity of the coordinates.

The above arguments apply also to the temporal abscissa, that, by
virtue of the \ref{eq:Evol_Durata}, may be considered as having as
conjugate momentum the Hamiltonian observable.

Effectively, a discrete structure of space and time at a Planck scale
may be considered not surprising. The point is so how to recover,
in a coherent manner, the canonical formalism. I think that this might
be done not in terms of relationship among observables but in terms
of expectation value and wave functions, for which continuity and
analyticity are recovered, even for finite Hilbert spaces.

This, however, will be the subject of future investigations.

\subsection{Time evolution and observation\label{subsec:Time-evolution-and-observation}}

A delicate point in the quantum mechanical description of the reality
is that of the relationship between time evolution and observation.
To be more precise, as already discussed in the second paper, an observation,
here meant as equivalent to a measurement process, breaks down the
continuity of the time evolution of the quantum states (it is here
convenient to assume the Schrödinger picture), causing a ``collapse''
of the density observable.

As widely discussed in the conclusions of the second paper, this way
of seeing the things is misleading: as a matter of fact, the density
observable is not a ``normal'' observable, but a property summarizing
the description that an observer can give of the physical system when
``adopting'' a suitable measurement setup, giving birth to a new
meta-observer. In this sense the density observable is relative to
the meta-observer, so each changing of the latter alters also the
first, through a logical - not physical - process of Bayesian inference.

But what does it happen if a second observer views ``externally''
the measurement act of the first? Here the word ``externally'' means
that there is no communication between the two observers. The situation
is well described by the \emph{relational vision} of the Quantum Mechanics
of Rovelli\citep{Rovelli1996,sep-qm-relational}.

To be clear, let's indicate with A (``Alice'') the first observer,
with B (``Bob'') the second one and with O the physical system ``measured''
by A.

From the point of view of Bob, the observer A is an \textbf{object}
of observation, describable as a quantum system entangled with O in
the measurement act. According to Bob the state of this entangled
system (A+O) evolves in a continuous manner over time - before, during
and after the measurement of A - according to the equations of motion
seen above (possibly using the von Neumann one). The final state is,
however, a \textbf{mixed} one, in which appears all the possible outcomes
of the measurement performed by A.

From the point of view of Alice - that doesn't know what B is seeing
and inferring - the measurement changes the observational situation,
due to the information acquired through the measurement setup, that
changes the meta-observer of which she is part. This fact forces a
change in the density observable adopted to describe O, that must
now correspond to the \textbf{pure} state associated with the outcome
of her measurement. But Alice cannot follow the time evolution of
the density observable during the measurement process, because she
cannot observe herself as an external object!

\textbf{If, after the measurement, Alice and Bob communicate with
each other, then they give rise to a new higher-level meta-observer,
so Bob ``observes'' a ``collapse'' of the density observable describing
the measurement process to agree with Alice for the outcome found}.
Again the ``collapse'' is due to the change of meta-observer, which
requires a revision in the density observable.

\subsection{Seven reasons to change}

Most of the results of this paper may not seem ``extraordinary'',
since, after all - apart from the treatment of transformations, the
continuity issues and the time reversal - they may be simply obtained
by standard textbook Quantum Mechanics by substitution of the operators
and the state vectors with the appropriate pseudo-observables - with
the remarkable exception of the time reversal. Effectively the principal
reason in doing so was just to show how easy is to switch from the
Dirac-von Neumann formalism to the new one.

So why would one have to change, if, after all, equations are, \textbf{formally},
the same?

In this conclusive subsection I will try to give a list of compelling
reasons to do this:
\begin{enumerate}
\item The standard formalism is based on a somewhat ``schizophrenic''
choice on the fundamental entities of the theory: abstract state vectors,
whose unclear and ambiguous ontology made flourish a plethora of different
interpretations of Quantum Mechanics, and an algebra of operators,
part of which corresponding to observable, whose definition is vague,
with a a far from being clear connection with the outcomes of a measurement.
The new formalism is characterized by an \textbf{ontology clear and
``parsimonious''}, making reference to the classification criteria
of Pykacz\citep{Pykacz2015}; gives the observables a due central
role, defining them as precise algebraic objects, and gives clear
and physically transparent definitions of the \textbf{only} type of
entities - the pseudo-observables - and of the operations and the
functionals needed to describe a physical system and to define the
measurement outcomes. The new theory is also ``embedded'' with a
clear and unambiguous interpretation, based both on the ``relational''
view of Rovelli and on Quantum Bayesianism, sweeping away all the
others from the field. 
\item Passing to new formalism requires almost no efforts, since \textbf{to
each valid expression in the Dirac-von Neumann formalism corresponds
a valid expression in the framework of the algebra of pseudo-observables,
simply substituting the operators and the state vectors with the corresponding
pseudo-observables}. So, for instance, if $\Psi_{i}$ and $\Psi_{j}$
are the state vector pseudo-observables corresponding to the ``kets''
$\ket i$ and $\ket j$ and if $A$ is the observable represented
by the operator $\hat{A}$, the matrix element $\bra i\hat{A}\ket j$
corresponds to the inner product $\inprod{\Psi_{i}}{A\,\Psi_{j}}$.
Therefore for a Physicist who embraces a purely \emph{instrumentalist}
vision\citep{Instrumentalism}, that is one - as a matter of fact
the majority\ldots{} - that explicitly avoids any explanatory role
of the theory, using it only as a mere instrument of calculation,
the changing is really minimal!
\item Even if, however, each expression in the Dirac-von Neumann formalism
has a corresponding one in the framework of the algebra of the pseudo-observables,
\textbf{the converse is not true!} So the latter offers new possibilities
to find more convenient, clear and compact demonstrations of relevant
physical relations.
\item The whole formalism of the algebra of the pseudo-observables is \textbf{simpler}
than the standard one, requiring only initial undergraduate level
knowledge, giving the opportunity to a \textbf{wider diffusion} of
a deeper understanding of Quantum Mechanics.
\item The theory was constructed on the basis of very general logical properties,
proving that quantum mechanics is the unique minimal description of
physical reality\citep{Piparo-I}. Unlike a diffuse practice, besides,
in this theory Mathematics is \textbf{constructed} to fit\textbf{
well-defined physical ideas}, not chosen among the existing models
to fit experimental data and generic physical conceptions.
\item The conceptual bases of the theory are physically transparent and
allows to easily overcome deep conceptual troubles in the Dirac-von
Neumann axiomatic system, as, for instance, the \textbf{measurement
problem}\citep{Piparo-II} (see also subsection \ref{subsec:Time-evolution-and-observation}).
\item \textbf{The algebra of the pseudo-observables entails the capability
of being a descriptive instrument that goes beyond the simply reformulation
of the Quantum Mechanics}. I have already obtained some preliminary
result which shows that this theory can give us a deeper insight in
the more intimate properties of the space and of the time, till the
Planck scale! This however will, hopefully, be the subject of future
investigations.
\end{enumerate}
\bibliographystyle{unsrt}
\addcontentsline{toc}{section}{\refname}\bibliography{../Bibliography}

\end{document}